\documentclass[twocolumn,pra,showpacs,floatfix,superscriptaddress,preprintnumbers]{revtex4}
\pdfoutput=1 

\usepackage{amsmath}
\usepackage{amsfonts}
\usepackage{amssymb}
\usepackage{graphicx}

\newcommand{\mathe}{\mathrm{e}}
\newcommand{\tmem}[1]{{\em #1\/}}
\newcommand{\tmmathbf}[1]{\ensuremath{\boldsymbol{#1}}}
\newcommand{\tmop}[1]{\ensuremath{\operatorname{#1}}}

\begin{document}

\title{Molecular Feshbach dissociation as a source for motionally entangled atoms}

\author{Clemens Gneiting}
\affiliation{Arnold Sommerfeld Center for Theoretical Physics, Ludwig-Maximilians-Universit{\"a}t M\"unchen, Theresienstra{\ss}e 37, 80333 Munich, Germany}
\author{Klaus Hornberger}
\affiliation{Arnold Sommerfeld Center for Theoretical Physics, Ludwig-Maximilians-Universit{\"a}t M\"unchen, Theresienstra{\ss}e 37, 80333 Munich, Germany}
\affiliation{Max Planck Institute for the Physics of Complex Systems, N\"othnitzer Stra{\ss}e 38, 01187 Dresden, Germany}

\begin{abstract}
  We describe the dissociation of a diatomic Feshbach molecule due to a
  time-varying external magnetic field in a realistic trap and guide setting.
  An analytic expression for the asymptotic state of the two ultracold atoms is
  derived, which can serve as a basis for the analysis of dissociation protocols
  to generate motionally entangled states. For instance, the gradual dissociation
  by sequences of magnetic field pulses may delocalize the atoms into macroscopically
  distinct wave packets, whose motional entanglement can be addressed interferometrically.
  The established relation between the applied magnetic field pulse and the generated
  dissociation state reveals that square-shaped magnetic field pulses minimize the momentum
  spread of the atoms. This is required to control the detrimental influence of dispersion
  in a recently proposed experiment to perform a Bell
  test in the motion of the two atoms [C.~Gneiting and K.~Hornberger, Phys. Rev. Lett. 101, 260503 (2008)].
\end{abstract}


\preprint{\textsf{published in Phys.~Rev.~A~{81}, 013423 (2010)}}
\maketitle

\section{Introduction}

The emerging field of ultracold atoms makes it possible to perform hitherto
unprecedented experiments on the quantum nature of material objects,
thus introducing a new quality compared to previous quantum experiments with
immaterial photons. This is owing to the experimental control over the atomic
state reaching a level where quantum mechanical effects become relevant. Given
that Bose-Einstein condensates (BECs) are nowadays routinely produced, they can
serve as an ideal starting point, for example, to probe condensed matter physics in a highly
controlled environment provided by optical lattices {\cite{Bloch2008a, Buchleitner2003a,witthaut:200402,hartung:020603}}.
A considerable extension of the scope of such experiments has been achieved by
exploiting Feshbach resonances to produce molecules and molecular BECs (mBECs) {\cite{Cubizolles2003a,Jochim2003a,Regal2003a,Greiner2003a,Strecker2003a,Herbig2003a,Xu2003a,Durr2004a}}.
The ability to control the interaction between the atoms with an external
magnetic field permits one to realize, for example, the BCS-BEC crossover
 {\cite{Bloch2008a}} or Efimov states \cite{Kraemer2006a}, and to establish coherent atom-molecule
oscillations {\cite{Donley2002a,Claussen2003a,Borca2003a,Duine2004a,Goral2005a}},
hinting at quantum-coherent chemistry.

Beyond that, it has been recognized that the controlled dissociation of mBECs
can also serve as a resource for entangled atom pairs that would permit the
demonstration of nonclassical correlations on a macroscopic level
{\cite{Kheruntsyan2005a,Kheruntsyan2006a}}. The use of such a controlled
dissociation for spectroscopic purposes has already been discussed theoretically \cite{Durr2005z,Hanna2006a} and demonstrated
experimentally {\cite{Mukaiyama2004a,Durr2004b}}. In a recent proposal, we
investigated the possibility of using an arranged dissociation of Feshbach
molecules in order to violate a Bell inequality in the motion of two atoms
{\cite{Gneiting2008a}}. There, a sequence of two short magnetic field pulses
dissociates a single molecule out of a dilute mBEC such that each atom is
delocalized into two macroscopically distinct wave packets propagating along
the laser guide. The associated dissociation-time entangled (DTE) state can be
considered the matter wave analog of a Bell state {\cite{Gneiting2009a}}, whose
capability to yield nonclassical correlations can be revealed by an interferometric
protocol reuniting the wave packets on each side \cite{Brendel1999a,Tittel2000a}. The violation of a Bell inequality,
however, imposes stringent requirements on the DTE state, which makes it
necessary to know the generated dissociation state in detail {\cite{Gneiting2009a}}.

The theory of Feshbach molecules is mainly concerned with an accurate
description of the molecules in the bound regime and its vicinity, where the
atoms interact strongly. For instance, it has been investigated in great detail that the
molecular state at constant magnetic field exhibits universal properties in the
vicinity of a Feshbach resonance, as well as the association and dissociation behaviors
under a linear magnetic field sweep, with emphasis on the converted fraction
(see {\cite{Hanna2006a,Kohler2006a,Nygaard2008a}} and Refs. therein). An appropriate description of the
situation proposed in {\cite{Gneiting2008a}}, however, requires a detailed
knowledge of the dissociation state for a more general temporal behavior of the
magnetic field, varying on short time scales compared to the inverse resonance width. 
While in the interaction regime of the atoms one is then forced to resort to 
numerical methods {\cite{Mies2000a}}, the restriction to the asymptotic situation
(i.e. for large interatomic distances and times long after the dissociation process)
comes with significant simplifications that permit one to evaluate
the generated dissociation state analytically.

In this article we present a coupled-channels formulation of the dissociation
process of an initially trapped Feshbach molecule exposed to a time-varying homogeneous
magnetic field. These single-molecule dynamics yield an adequate description in the
relevant case of the dissociation of a single molecule out of a dilute mBEC, where
interactions between the molecules and statistical effects do not play a role.
The formulation includes the center of mass motion required for the
complete description of the two-particle system. The dissociation is considered to
take place in a shallow trap from where the dissociated atoms are injected
into a strong guiding potential, which confines them to the longitudinal motion
along the guide axis. Based on the time-dependent coupled-channel equations,
we derive an analytic expression for the asymptotic dissociation state of the
ultracold atoms, allowing us to investigate the relation between the form
of the applied magnetic field pulse and the resulting two-particle dissociation
state. While Gaussian-shaped dissociation pulses turn out to yield a rather
bulky momentum distribution of the two-particle state, we demonstrate that idealized
square-shaped magnetic field pulses optimize the momentum distribution with respect to
its sharpness.

The structure of this article is as follows. In Section \ref{ExperimentalSetup} we
describe the assumed geometry of trap and guiding potentials, which is the natural
configuration for dissociation experiments. The time-dependent coupled-channel equations
are formulated in Section \ref{SectionCoupledChannelsEquations}, and then reduced to an
integro-differential equation for the closed-channel amplitude and an associated
equation for the background channel state. Section \ref{AsymptoticDissociationState}
shows how the asymptotic dissociation state can be expressed in terms of the
Fourier transform of the closed-channel amplitude. The approximate dynamics of the
latter for a given shape of the magnetic field pulse is described in
Section \ref{ClosedChannelAmplitudeDynamics}; this allows us, in Section \ref{OptimalPulse},
to discuss the optimal form of the pulse if a narrow momentum distribution of the
dissociation products is required, such as for the test of the Bell inequality
proposed in \cite{Gneiting2008a}. Applications of the Feshbach dissociation scheme
as a resource of entangled atom states beyond the scenario in {\cite{Gneiting2008a}}
are discussed in the Conclusions.

\section{\label{ExperimentalSetup}Trap and guide configuration}

Before presenting the coupled-channels formalism, we outline an experimental
setup that meets the conditions assumed in our dissociation scenario. An
explicit quantitative elaboration was given in {\cite{Gneiting2009b}}.

\begin{figure}[tb]
  \includegraphics[width=\columnwidth]{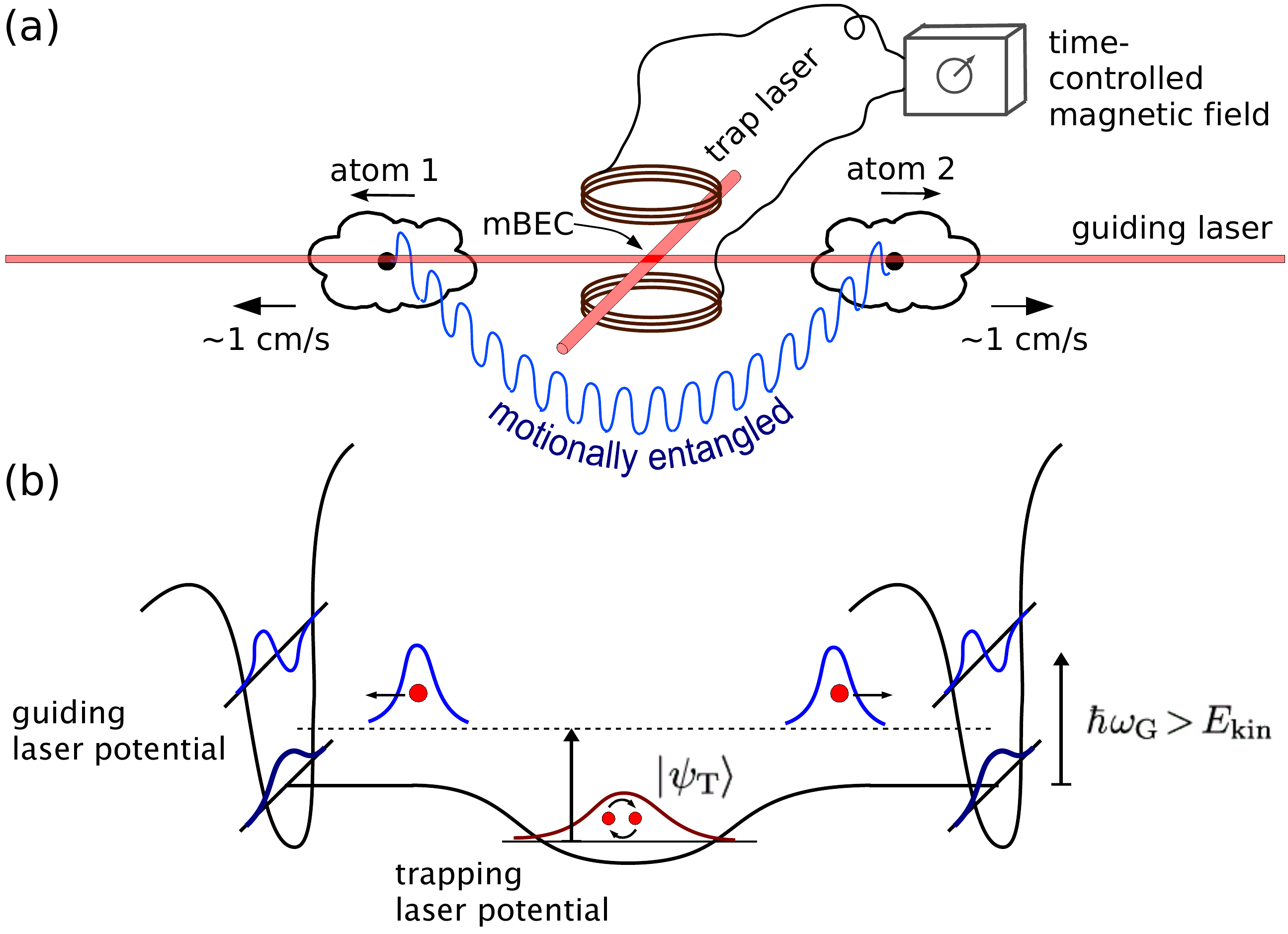}
  \caption{\label{Setting}(color online) a) Setup for generating pairs of
  motionally entangled atoms by the Feshbach dissociation of molecules.
  Initially, a BEC of on the order of $10^2$ Feshbach molecules 
  resides in a dipole trap produced by two 
  crossing laser beams. An externally controlled, homogeneous magnetic field induces
  the dissociation of one molecule per trial and thus generates a pair
  of atoms moving along the laser guide at a velocity on
  the order of 1 cm/s. The asymptotic two-atom state in the wave guide is
  determined by the trap and guide geometry and by the shape of the
  dissociation pulse. b) The dissociation pulse promotes the trapped
  molecule to a pair of counter-propagating atoms. The laser parameters are chosen such
  that the energy supply from the magnetic field sweep exceeds the trap laser potential,
  while it cannot lead to an additional transversal excitation if the kinetic energy of
  the atoms $E_{\text{kin}}$ remains below the harmonic energy gap,
  $\hbar \omega_{\text{G}} > E_{\tmop{kin}}$.}
\end{figure}

Consider a BEC of Feshbach molecules in an optical dipole trap, which can be established by
two perpendicularly crossing laser beams, see Figure \ref{Setting} a). The weak
trap laser guarantees longitudinal confinement within the wave guide produced by the
strong guiding laser. We take the BEC, of the order of $10^2$ molecules, to be sufficiently
dilute such that one may neglect interactions between different molecules. This may
be accomplished by choosing a mBEC of fermionic constituents, for example $^6 \tmop{Li}_2$,
where Pauli blocking further reduces the effect of intermolecular interactions and thus
enhances the lifetime of the mBEC \cite{Jochim2003aORIG}.

The molecules can then be considered to be in a product state with the center of mass
motion given by the ground state of the trap, whereas the relative motion is in a
bound molecular state. The latter can be turned into a Feshbach resonance by varying
the external magnetic field, allowing one to dissociate
the atoms in a controlled way. By applying one or several appropriately chosen
dissociation pulses, a single molecule dissociates into two
counter-propagating atoms on average, and we postselect the single-dissociation events.

The pulses provide the atoms with a kinetic energy sufficiently large to overcome
the trap potential in longitudinal direction, but still below the threshold to
get beyond the ground state transversally, see Figure \ref{Setting} b). This
way we may end up with two dissociated atoms, counter-propagating with a velocity on
the order of 1 cm/s along the guiding laser axis, whose two-particle
state is determined by the initial state of the molecule and the dissociation
pulse shape. By applying sequences of dissociation pulses, one can design
highly nonclassical, motionally entangled states that can further be processed for various
fundamental tests of quantum mechanics, such as to violate a Bell inequality
{\cite{Gneiting2008a}}.

\section{\label{SectionCoupledChannelsEquations}coupled-channels formulation}

The dynamics of Feshbach molecules is appropriately described by the coupled
channels formulation. In our case, the channels are characterized by different
nuclear spin configurations of the two-atom system. The different magnetic
field dependences of the corresponding energy levels make it possible to
manipulate the system externally. In the following we adopt the notation and
the conventions from {\cite{Kohler2006a}} as far as possible. As a novelty,
we must include also the center of mass motion of the two atoms.
We assume that the magnetic field remains always
in the vicinity of a single Feshbach resonance, allowing us to restrict
the description to two channels: the closed channel of the
energetically more favorable spin configuration supporting
the molecular bound state, and the background channel, where
the dissociated atoms are asymptotically free.

\subsection{Hamiltonian and coupled-channel equations}

The two-channel Hamiltonian for the two atoms (here taken to be of equal
mass $m$) can be written as
\begin{eqnarray}
  H_{\tmop{tot}}  & = & H_{\tmop{cl}} | \tmop{cl} \rangle \langle \tmop{cl} |
  + H_{\tmop{bg}} | \tmop{bg} \rangle \langle \tmop{bg} |\nonumber\\
  &  & + W | \tmop{bg} \rangle \langle \tmop{cl} | + W^{\dagger} | \tmop{cl}
  \rangle \langle \tmop{bg} |,
\end{eqnarray}
with the closed-channel Hamiltonian
\begin{eqnarray}
  H_{\tmop{cl}} & = & - \frac{\hbar^2}{2 m} \nabla_1^2 - \frac{\hbar^2}{2 m}
  \nabla^2_2 + V_{\tmop{cl}} (|\tmmathbf{x}_1 -\tmmathbf{x}_2 |, B (t)) \nonumber \\
  &  & + V_{\text{T}} (\tmmathbf{x}_1) + V_{\text{T}} (\tmmathbf{x}_2)
  + V_{\text{G}} (\tmmathbf{x}_1) + V_{\text{G}} (\tmmathbf{x}_2)
\end{eqnarray}
and the background (open) channel Hamiltonian
\begin{eqnarray}
  H_{\tmop{bg}} & = & - \frac{\hbar^2}{2 m} \nabla_1^2 - \frac{\hbar^2}{2 m}
  \nabla^2_2 + V_{\tmop{bg}} (|\tmmathbf{x}_1 -\tmmathbf{x}_2 |) \nonumber \\
  &  & + V_{\text{T}} (\tmmathbf{x}_1) + V_{\text{T}} (\tmmathbf{x}_2)
  + V_{\text{G}} (\tmmathbf{x}_1) + V_{\text{G}} (\tmmathbf{x}_2) .
\end{eqnarray}
Here $V_{\text{T}} (\tmmathbf{x}_i)$ and $V_{\text{G}} (\tmmathbf{x}_i)$
denote the trapping and guiding laser potential, respectively ($V_{\text{G}}$
may contain a linear shift due to the gravitational potential).
$V_{\tmop{cl}} (|\tmmathbf{x}_1 -\tmmathbf{x}_2 |, B (t))$ and $V_{\tmop{bg}}
(|\tmmathbf{x}_1 -\tmmathbf{x}_2 |)$, on the other hand, describe the
interatomic potentials for each channel. In general, these potentials differ
for the different channels, reflecting their dependence on the spin
configuration. In the chosen convention, the zero of total energy is defined in
absence of the laser potentials by the background channel dissociation
threshold with the center of mass at rest. Then only the closed-channel potential
$V_{\tmop{cl}} (|\tmmathbf{x}_1 -\tmmathbf{x}_2 |, B (t))$ depends on the
external magnetic field $B (t)$, describing an overall shift with respect to
the background channel dissociation threshold.

The off-diagonal elements $W$ denote the energies associated with the spin exchange
interaction and provide the inter-channel coupling. We assume $W$ to be diagonal in
position (i.e. independent of momentum) and to depend only on the interatomic distance
$|\tmmathbf{x}_1-\tmmathbf{x}_2|$ from now on. For the following it is
useful to reformulate the Hamiltonian in center of mass (cm) and relative
(rel) coordinates, $\tmmathbf{x}_{\tmop{cm}} = (\tmmathbf{x}_1
+\tmmathbf{x}_2) / 2$ and $\tmmathbf{x}_{\tmop{rel}} =\tmmathbf{x}_1
-\tmmathbf{x}_2$, respectively, with total mass $M = 2 m$ and reduced mass $\mu = m / 2$.
The closed-channel Hamiltonian thus reads as
\begin{eqnarray}
  H_{\tmop{cl}} & = & - \frac{\hbar^2}{2 M} \nabla^2_{\tmop{cm}} -
  \frac{\hbar^2}{2 \mu} \nabla^2_{\tmop{rel}} + V_{\tmop{cl}}
  (|\tmmathbf{x}_{\tmop{rel}} |, B (t)) \nonumber \\
  &  & + V_{\text{T}} (\tmmathbf{x}_{\tmop{cm}} +\tmmathbf{x}_{\tmop{rel}} /
  2) + V_{\text{T}} (\tmmathbf{x}_{\tmop{cm}} -\tmmathbf{x}_{\tmop{rel}} /
  2)\\
  &  & + V_{\text{G}} (\tmmathbf{x}_{\tmop{cm}} +\tmmathbf{x}_{\tmop{rel}} /
  2) + V_{\text{G}} (\tmmathbf{x}_{\tmop{cm}} -\tmmathbf{x}_{\tmop{rel}} / 2), \nonumber
\end{eqnarray}
and similar for $H_{\text{bg}}$ with $V_{\tmop{cl}} (|\tmmathbf{x}_{\tmop{rel}} |, B (t))$
replaced by $V_{\tmop{bg}}(|\tmmathbf{x}_{\tmop{rel}}|)$. Note that the center of
mass motion is only indirectly affected by the homogeneous external magnetic
field $B$, due to the presence of the trapping potentials. Writing
\begin{equation}
 | \Psi_{\tmop{tot}} (t) \rangle = \Phi_{\tmop{cl}}
   (\tmmathbf{x}_{\tmop{cm}}, \tmmathbf{x}_{\tmop{rel}}, t) | \tmop{cl}
   \rangle + \Phi_{\tmop{bg}} (\tmmathbf{x}_{\tmop{cm}},
   \tmmathbf{x}_{\tmop{rel}}, t) | \tmop{bg} \rangle, 
\end{equation}
we can infer from the time-dependent Schr\"odinger equation the coupled
channel equations
\begin{eqnarray}
  i \hbar | \Phi_{\tmop{cl}} (t) \rangle & = & H_{\tmop{cl}} (B (t)) |
  \Phi_{\tmop{cl}} (t) \rangle + W | \Phi_{\tmop{bg}} (t) \rangle 
  \label{CoupledChannelsEquations}\\
  i \hbar | \Phi_{\tmop{bg}} (t) \rangle & = & H_{\tmop{bg}} |
  \Phi_{\tmop{bg}} (t) \rangle + W | \Phi_{\tmop{cl}} (t) \rangle . \nonumber
\end{eqnarray}
The closed channel and background channel two-particle state components
$| \Phi_{\tmop{cl}} (t) \rangle$ and $| \Phi_{\tmop{bg}} (t) \rangle$ are thus
not normalized in general. It is reasonable to assume the transverse motion
of the atoms to be constrained to the region of validity of the harmonic
approximation of the guiding laser potential, and the harmonic approximation
to be applicable also for the trap laser potential in the closed channel. We
then can write (adopting cylindrical coordinates)
\begin{eqnarray}
  V_{\text{G}} (\tmmathbf{x}_1) + V_{\text{G}} (\tmmathbf{x}_2) & = & -U_{0,
  \text{G}} + \frac{m}{2} \omega^2_{\text{G}} \rho^2_1 - U_{0, \text{G}} +
  \frac{m}{2} \omega^2_{\text{G}} \rho^2_2 \nonumber\\
  V_{\text{T}} (\tmmathbf{x}_1) + V_{\text{T}} (\tmmathbf{x}_2) & = & -U_{0,
  \text{T}} + \frac{m}{2} \omega^2_{\text{T}} z^2_1 - U_{0, \text{T}} +
  \frac{m}{2} \omega^2_{\text{T}} z^2_2 . \nonumber \\&&
\end{eqnarray}
The harmonic approximation comes with the virtue not to couple the center of
mass and relative motion,
\begin{eqnarray}
  V_{\text{G}} (\tmmathbf{x}_1) + V_{\text{G}} (\tmmathbf{x}_2) & = & -2 \, U_{0,
  \text{G}} + \frac{M}{2} \omega^2_{\text{G}} \rho^2_{\tmop{cm}}
  + \frac{\mu}{2} \omega^2_{\text{G}} \rho^2_{\tmop{rel}} \nonumber \\
  V_{\text{T}} (\tmmathbf{x}_1) + V_{\text{T}} (\tmmathbf{x}_2) & = & -2 \, U_{0,
  \text{T}} + \frac{M}{2} \omega^2_{\text{T}} z^2_{\tmop{cm}}
  + \frac{\mu}{2} \omega^2_{\text{T}} z^2_{\text{rel}} . \nonumber\\&&
\end{eqnarray}
The initial bound state can therefore be taken to be separable with respect to
its center of mass and relative motion. Of course, the harmonic approximation
for the trap laser potential breaks down when the dissociated atoms leave the
trap. Then, the center of mass motion ceases to be bound by the trap potential
but rather undergoes a free propagation resulting in a dispersive broadening
on the spot. So, even though initially only the relative motion is affected by
the external magnetic field, its effective coupling in the background channel
to the center of mass also couples the motion of the latter indirectly to the
external magnetic field.

\subsection{Single-resonance approximation}

It is legitimate {\cite{Kohler2006a}} to take the relative motion of the closed
channel state component to be proportional to the underlying bare
resonance state $| \phi_{\tmop{res}} \rangle$, which is defined by
\begin{multline}
   \left[ - \frac{\hbar^2}{2 \mu} \nabla_{\tmop{rel}}^2 + V_{\tmop{cl}}
   (|\tmmathbf{x}_{\tmop{rel}} |, B (t)) \right] \phi_{\tmop{res}}
   (\tmmathbf{x}_{\tmop{rel}}) 
\\
= E_{\tmop{res}} (B (t)) \phi_{\tmop{res}}
   (\tmmathbf{x}_{\tmop{rel}}) .
\end{multline}
Note that the time-dependent external magnetic field affects only its energy,
which is taken to vanish at the resonance, $E_{\tmop{res}} (B_{\tmop{res}}) = 0$,
such that $E_{\tmop{res}}$ describes the energetic offset of $|\phi_{\tmop{res}}
\rangle$ from the background channel dissociation threshold.

Since the laser potentials vary weakly over the spatial extent of $| \phi_{\tmop{res}} \rangle$,
this resonance state remains a valid approximation even in the presence of the trap.
We assume that $| \phi_{\tmop{res}} \rangle$ is spherically symmetric and thus
supports an s-wave resonance. If we further take into account that the
center of mass motion of the closed channel is completely determined by
the longitudinal and transversal trap ground states
$|\psi_{\text{T}} \rangle$ and $| \varphi^{\tmop{cm}}_{0, 0} \rangle$,
respectively, we can write
\begin{equation}
  \label{SingleResonanceApproximation} \Phi_{\tmop{cl}}
  (\tmmathbf{x}_{\tmop{cm}}, \tmmathbf{x}_{\tmop{rel}}, t) = C (t)
  \psi_{\text{T}} (z_{\tmop{cm}}) \varphi^{\tmop{cm}}_{0, 0}
  (\rho_{\tmop{cm}}) \phi_{\tmop{res}} (\tmmathbf{x}_{\tmop{rel}}),
\end{equation}
where
\begin{multline}
  [- \frac{\hbar^2}{2 M} \nabla_{\tmop{cm}}^2 + 2 V_{\text{T}}
  (z_{\tmop{cm}})  + 2 V_{\text{G}} (\rho_{\tmop{cm}})] \psi_{\text{T}}
  (z_{\tmop{cm}}) \varphi^{\tmop{cm}}_{0, 0} (\rho_{\tmop{cm}})  \\
= [-2 U_{0, \text{T}} + \hbar \omega_{\text{T}} / 2
  - 2 U_{0, \text{G}} + \hbar \omega_{\text{G}}] \psi_{\text{T}}
  (z_{\tmop{cm}}) \varphi^{\tmop{cm}}_{0, 0} (\rho_{\tmop{cm}}) .
\end{multline}
In the single-resonance approximation (extended by the trapped center of mass
motion) (\ref{SingleResonanceApproximation}), the spatial shape of the closed
channel state component is not affected by the external magnetic field and
therefore time independent. The closed-channel amplitude $C (t)$ therefore
captures the complete effect of the time-varying magnetic field. Using the
single-resonance approximation (\ref{SingleResonanceApproximation})
and introducing the abbreviation
$U_{\tmop{cl}} = -2 U_{0, \text{T}} + \hbar \omega_{\text{T}} / 2 - 2 U_{0,
\text{G}} + \hbar \omega_{\text{G}}$, we can thus rewrite the coupled-channels
equations (\ref{CoupledChannelsEquations}) as
\begin{eqnarray}
  \label{ClosedChannelEquation}
  i \hbar \partial_t C (t) & = & [E_{\tmop{res}} (B (t)) + U_{\tmop{cl}}] C
  (t)\\&& + \langle \psi_{\text{T}} | \langle \varphi^{\tmop{cm}}_{0, 0} | \langle
  \phi_{\tmop{res}} |W| \Phi_{\tmop{bg}} (t) \rangle \nonumber
\end{eqnarray}
\begin{eqnarray}
  (i \hbar \partial_t - H_{\tmop{bg}}) | \Phi_{\tmop{bg}} (t) \rangle & = & C
  (t) W | \psi_{\text{T}} \rangle | \varphi^{\tmop{cm}}_{0, 0} \rangle |
  \phi_{\tmop{res}} \rangle .\quad  \label{BackgroundChannelEquation}
\end{eqnarray}

\subsection{Formal Green's solution}

Interpreting the right-hand side of (\ref{BackgroundChannelEquation}) as a source term for
the background channel state component $| \Phi_{\tmop{bg}} (t) \rangle$
suggests to solve (\ref{BackgroundChannelEquation}) formally using the Green's
function of the background channel $G_{\tmop{bg}} (t, t')$, which satisfies
\begin{equation}
 (i \hbar \partial_t - H_{\tmop{bg}}) G_{\tmop{bg}} (t, t') = \delta (t -
   t') . 
\end{equation}
If we further make use of the connection between the (retarded) Green's
function and the time evolution operator,
$ G_{\tmop{bg}} (t, t') =  U_{\tmop{bg}} (t, t') \Theta (t -
   t')/({i \hbar})$,
we can write the background channel state component as
\begin{equation}
  \label{BackgroundChannelFormalSolution} | \Phi_{\tmop{bg}} (t) \rangle =
  \frac{1}{i \hbar}  \int^t_{- \infty} \text{d} t' C (t') U_{\tmop{bg}} (t,
  t') W | \psi_{\text{T}} \rangle | \varphi^{\tmop{cm}}_{0, 0} \rangle |
  \phi_{\tmop{res}} \rangle .
\end{equation}
In our scenario, the boundary conditions prohibit a homogeneous solution.
Physically, this reflects the fact that the closed channel is the only source
for the background channel, in particular there are no further sources at
infinity (for example incoming and scattered particles). A closed equation for the
closed-channel amplitude $C (t)$ arises from inserting the formal solution
(\ref{BackgroundChannelFormalSolution}) into (\ref{ClosedChannelEquation}),
which yields
\begin{multline}
  \label{ClosedChannelDecoupled} [i \hbar \partial_t - E_{\tmop{res}} (B (t))
  - U_{\tmop{cl}}] C (t) = \frac{1}{i\hbar} \int^t_{- \infty} \text{d} t' C (t')
\\\times \langle
  \psi_{\text{T}} | \langle \varphi^{\tmop{cm}}_{0, 0} | \langle
  \phi_{\tmop{res}} |W U_{\tmop{bg}} (t, t') W| \psi_{\text{T}} \rangle |
  \varphi^{\tmop{cm}}_{0, 0} \rangle | \phi_{\tmop{res}} \rangle .
\end{multline}
With (\ref{BackgroundChannelFormalSolution}) and
(\ref{ClosedChannelDecoupled}) we arrived at a decoupled set of
equations that divides the determination of the background channel
dissociation state into two parts, first solving
(\ref{ClosedChannelDecoupled}) for $C (t)$, then using the solution in
(\ref{BackgroundChannelFormalSolution}) for the calculation of $|
\Phi_{\tmop{bg}} \rangle$. The dynamics of the closed-channel amplitude $C
(t)$ is explicitly driven by the external magnetic field $B (t)$, reflected in
the left-hand side of (\ref{ClosedChannelDecoupled}). The bare background channel
Hamiltonian $H_{\tmop{bg}}$, on the other hand, does not depend on the
external magnetic field, which will allow us in Section
\ref{AsymptoticDissociationState} to expand the right-hand side of
(\ref{BackgroundChannelFormalSolution}) in terms of its (time- and
coupling-independent) energy eigenfunctions. The right-hand side of
(\ref{ClosedChannelDecoupled}) describes the back action of the background
channel state component on the dynamics of $C (t)$ due to the coupling $W$. A
solution to (\ref{ClosedChannelDecoupled}) will be given in Section
\ref{ClosedChannelAmplitudeDynamics}.

As a last remark, we note that it might seem suggestive to first solve the
time-independent coupled-channel equations for a stationary magnetic field $B$,
and then to take the corresponding static decay rate to describe the dissociation
in the time-dependent case $B(t)$, as it was done in \cite{Mukaiyama2004a}.
However, we will see later in this article that this quasi-stationary approach is not sufficient
for our purposes.

\section{\label{AsymptoticDissociationState}Asymptotic dissociation state}

The formal expression (\ref{BackgroundChannelFormalSolution}) describes the
background channel state component in full generality. In the scenario
described in Section \ref{ExperimentalSetup}, however, we only need to know
the dissociation state for large interatomic distances and for times long
after the dissociation process. Moreover, the relevant dissociation states
are sharply peaked in the ultracold regime, in the sense that the width of
the momentum distribution is much smaller than its average momentum, because
only then they are useful with respect to further employment such as
the Bell test in the motion {\cite{Gneiting2008a}}. The above restrictions
admit significant simplifications that permit us to provide an analytic
expression for the dissociation state in the asymptotic regime.

\subsection{Large time limit}

As a first step, we expand the background channel time evolution operator
$U_{\tmop{bg}} (t, t')$ in an appropriate energy eigenbasis of the background
channel Hamiltonian.
\begin{equation}
  \label{EvolutionOperatorInEnergyBasis} U_{\tmop{bg}} (t, t') = \mathe^{- i
  H_{\tmop{bg}} (t - t')} = \sum_E \mathe^{- i E (t - t') / \hbar} |E
  \rangle_{\tmop{bg}} \langle E|_{\tmop{bg}},
\end{equation}
where $H_{\tmop{bg}} |E \rangle_{\tmop{bg}} = E |E \rangle_{\tmop{bg}}$;
adequate quantum numbers for our setup will be specified later in this article. Note that
the involved vectors are two-particle states. As mentioned above, this
representation is only possible for the bare background Hamiltonian, which
does not depend on the external magnetic field. Since at large interatomic
distances only the continuum states survive, we can drop the bound states in
(\ref{BackgroundChannelFormalSolution}) for large $|\tmmathbf{x}_{\text{rel}}|$ and get
\begin{align}
   \langle \tmmathbf{x}_{\text{cm}} , \tmmathbf{x}_{\text{rel}} | \Phi_{\tmop{bg}} (t) \rangle
   & \underset{|\tmmathbf{x}_{\text{rel}}| \rightarrow \infty}{\sim} 
   \frac{1}{i \hbar}  \sum_{E > U_{\text{bg}}} \int^t_{- \infty} \text{d} t' C (t')
   \\&\quad\quad\times\mathe^{- i E (t - t') / \hbar} 
\langle \tmmathbf{x}_{\text{cm}} , \tmmathbf{x}_{\text{rel}} |E \rangle_{\tmop{bg}}
 \nonumber \\
    & \quad\quad\times \langle E|_{\tmop{bg}} W | \psi_{\text{T}} \rangle |
   \varphi^{\tmop{cm}}_{0, 0} \rangle | \phi_{\tmop{res}} \rangle .\nonumber
\end{align} 
The sum over the energy eigenstates takes into account that the zero of energy is defined in absence
of the confining lasers, whose presence shifts the (longitudinal) continuum threshold by an offset
$U_{\text{bg}} = -2 U_{0, \text{G}} + 2 \hbar \omega_{\text{G}}$. We arrange the dissociation such
that after its completion the magnetic field persists at a base value $B_0$ below the dissociation
threshold. The closed-channel amplitude then shows a simple time dependence in the large time regime,
$C(t) = C_0 \text{exp}(-i E_0 t / \hbar)$, such that for times long after the completion we can
replace the upper integration boundary by infinity without modifying the integral for energies
above the dissociation threshold. This allows us to interpret the integration over $t'$ as the Fourier
transform $\tilde{C} (\omega) = \int^{\infty}_{- \infty} \text{d} t \, \text{exp}(i \omega t) C (t)$
of the closed-channel amplitude $C (t)$, yielding
\begin{align}
   \langle \tmmathbf{x}_{\text{cm}} , \tmmathbf{x}_{\text{rel}} | \Phi_{\tmop{bg}} (t) \rangle
   & \underset{t \rightarrow \infty}{\underset{|\tmmathbf{x}_{\text{rel}}| \rightarrow \infty}{\approx}} 
   \frac{1}{i \hbar}  \sum_{E > U_{\text{bg}}} \mathe^{- i E t / \hbar}  \tilde{C}
   (E / \hbar)\nonumber\\&\quad\quad\times \langle E|_{\tmop{bg}} W | \psi_{\text{T}} \rangle |
   \varphi^{\tmop{cm}}_{0, 0} \rangle | \phi_{\tmop{res}} \rangle \nonumber \\
   & \quad\quad \times \langle \tmmathbf{x}_{\text{cm}} , \tmmathbf{x}_{\text{rel}} |E \rangle_{\tmop{bg}} .
\end{align}
We thus find that the asymptotic dissociation state can be interpreted as evolving
in the background channel with the initial state in energy representation given by
$\tilde{C} (E / \hbar) \langle E|_{\tmop{bg}} W | \psi_{\text{T}} \rangle |
\varphi^{\tmop{cm}}_{0, 0} \rangle | \phi_{\tmop{res}} \rangle$. We will see
that this expression describes two counter-propagating atoms with well-defined
momenta if $\tilde{C} (\omega)$ is peaked at an energy in the ultracold regime.
The corresponding wave functions then have de Broglie wavelengths and spatial
extensions on the order of micro- to millimeters, all being features desired
for applications such as further interferometric manipulation.

\subsection{Quantum numbers}

Note that the energy eigenvalues in (\ref{EvolutionOperatorInEnergyBasis})
are highly degenerate; in order to single out a unique energy basis, we choose as
asymptotically well-defined, commuting observables the complete set
$\hat{p}_{\tmop{cm}}$, $\hat{p}_{\tmop{rel}}$ and $H_{\bot}$, where $H_{\bot}$
denotes the transversal Hamiltonian. Assuming harmonic transversal confinement,
we can write
\begin{align}
  \sum_{E > U_{\text{bg}}}& |E \rangle_{\tmop{bg}} \langle E|_{\tmop{bg}}  \nonumber\\
= &
  \underset{n^{\tmop{rel}}_{\text{G}}, m^{\tmop{rel}}_{\text{G}}}{\sum_{n^{\tmop{cm}}_{\text{G}},
  m^{\tmop{cm}}_{\text{G}}}}  \int^{\infty}_{- \infty} \text{d} p_{\tmop{cm}}
  \int^{\infty}_{- \infty} \text{d} p_{\tmop{rel}}
\nonumber\\&\times |p_{\tmop{cm}},
  n^{\tmop{cm}}_{\text{G}}, m^{\tmop{cm}}_{\text{G}}, p_{\tmop{rel}},
  n^{\tmop{rel}}_{\text{G}}, m^{\tmop{rel}}_{\text{G}} \rangle_{\tmop{bg}} \nonumber \\
  &  \times \langle p_{\tmop{cm}}, n^{\tmop{cm}}_{\text{G}}, m^{\tmop{cm}}_{\text{G}},
  p_{\tmop{rel}}, n^{\tmop{rel}}_{\text{G}}, m^{\tmop{rel}}_{\text{G}}
  |_{\tmop{bg}} . \label{AsymptoticQuantumNumbers}
\end{align}
Here $n_{\text{G}} = 0,1,\dots$ denotes the radial occupation number and
$m_{\text{G}} = -n_{\text{G}},\dots,n_{\text{G}}$ the corresponding angular momentum.
Since we assume that $W| \phi_{\tmop{res}} \rangle$ is spherically symmetric ($s$-wave resonance),
$W| \psi_{\text{T}} \rangle | \varphi^{\tmop{cm}}_{0, 0} \rangle | \phi_{\tmop{res}} \rangle$ is
both cylindrically symmetric and symmetric under exchange of the two particles, hence only the
states $|p_{\tmop{cm}}, n, 0, p_{\tmop{rel}}, n, 0 \rangle$ sharing these symmetries can be
occupied.

If we take $\tilde{C} (E / \hbar)$ to be peaked at a sufficiently low energy and if the laser
guide is chosen appropriately, we can ensure that only the transversal ground state is
energetically accessible. In this single-mode regime we have
\begin{align}
  \langle \tmmathbf{x}_{\text{cm}} , \tmmathbf{x}_{\text{rel}} | & \Phi_{\tmop{bg}} (t) \rangle \nonumber \\
  \underset{t \rightarrow \infty}{\underset{|\tmmathbf{x}_{\text{rel}}| \rightarrow \infty}{\approx}}
  & \frac{1}{i \hbar}  \int^{\infty}_{- \infty} \text{d} p_{\tmop{cm}} \int^{\infty}_{- \infty} \text{d} p_{\tmop{rel}}
  \, \mathe^{- i (U_{\tmop{bg}} + p^2_{\tmop{cm}} / 2 M + p^2_{\tmop{rel}} / 2 \mu) t / \hbar}
\nonumber\\&\times\tilde{C} \left( \frac{U_{\tmop{bg}} + p^2_{\tmop{cm}} / 2 M + p^2_{\tmop{rel}} / 2 \mu}{\hbar} \right) \nonumber \\
  & \times \langle p_{\tmop{cm}}, 0, 0, p_{\tmop{rel}}, 0, 0|_{\tmop{bg}} W| \psi_{\text{T}}
  \rangle | \varphi^{\tmop{cm}}_{0, 0} \rangle | \phi_{\tmop{res}} \rangle
\nonumber\\&\times
  \langle \tmmathbf{x}_{\text{cm}} , \tmmathbf{x}_{\text{rel}}
  |p_{\tmop{cm}}, 0, 0, p_{\tmop{rel}}, 0, 0 \rangle_{\tmop{bg}}.
\end{align}
Even taking $\tilde{C} (E / \hbar)$ to be sharply peaked in energy, it still admits
the whole class of degenerate eigenstates that fall into that energetic region. As we
show next, the matrix element $\langle p_{\tmop{cm}}, 0, 0, p_{\tmop{rel}}, 0,
0|_{\tmop{bg}} W| \psi_{\text{T}} \rangle | \varphi^{\tmop{cm}}_{0, 0} \rangle
| \phi_{\tmop{res}} \rangle$ effects a further restriction of the accessible
eigenstates, yielding the physically appropriate description of the situation.

\subsection{Asymptotic center of mass motion}

The main effect of the longitudinal trap is to correlate the motion of the interatomic
distance with the center of mass, since the center of mass evolution is confined
for small interatomic distances, while it is free for distances beyond the size
of the trap. For a sufficiently narrow energy distribution this merely results in
a global time delay for the start of the free center of mass propagation, which
may be neglected at the time scales of the asymptotic regime. The longitudinal center
of mass motion can thus be described by the momentum eigenstates $|p_{\text{cm}} \rangle$,
that is by
\[ |p_{\tmop{cm}}, 0, 0, p_{\tmop{rel}}, 0, 0 \rangle_{\tmop{bg}} \approx
   |p_{\tmop{cm}} \rangle | \varphi^{\tmop{cm}}_{0, 0} \rangle
   |p_{\tmop{rel}}, 0, 0 \rangle_{\tmop{bg}} . \]
Here $|p_{\tmop{rel}}, 0, 0 \rangle_{\tmop{bg}}$ denotes the eigenstates of the
Hamiltonian for the relative motion, which still contains the traps and the
interatomic potential $V_{\tmop{bg}}$. The dissociation state then reads
\begin{align}
  \langle \tmmathbf{x}_{\text{cm}} , \tmmathbf{x}_{\text{rel}} | \Phi_{\tmop{bg}} (t) \rangle
  & \underset{t \rightarrow \infty}{\underset{|\tmmathbf{x}_{\text{rel}}| \rightarrow \infty}{\approx}} 
  \frac{1}{i \hbar}  \int^{\infty}_{- \infty} \text{d} p_{\tmop{cm}} \int^{\infty}_{- \infty} \text{d} p_{\tmop{rel}}
  \nonumber\\&\quad\quad
\times\mathe^{- i (U_{\tmop{bg}} + p^2_{\tmop{cm}} / 2 M + p^2_{\tmop{rel}} / 2 \mu) t / \hbar} \nonumber \\
  & \quad\quad  \times \tilde{C} \left( \frac{U_{\tmop{bg}} + p^2_{\tmop{cm}} / 2 M + p^2_{\tmop{rel}} / 2 \mu}{\hbar} \right)
  \nonumber\\&\quad\quad
\times  \langle p_{\tmop{cm}} | \psi_{\text{T}} \rangle \langle p_{\tmop{rel}}, 0, 0|_{\tmop{bg}} W| \phi_{\tmop{res}} \rangle \nonumber \\
  &  \quad\quad \times \langle z_{\text{cm}} |p_{\tmop{cm}} \rangle \langle \rho_{\text{cm}} | \varphi^{\tmop{cm}}_{0, 0} \rangle
  \nonumber\\&\quad\quad\times
  \langle \tmmathbf{x}_{\text{rel}} |p_{\tmop{rel}}, 0, 0 \rangle_{\tmop{bg}} .
\end{align}
The matrix element $\langle p_{\tmop{cm}} | \psi_{\text{T}} \rangle$, given
by the longitudinal harmonic trap ground state, 
guarantees that the center of mass motion remains
centered at vanishing momentum, as required by momentum conservation in the
dissociation process. Since we take the energy to be
in the ultracold regime of the background channel potential $V_{\tmop{bg}}$,
the matrix element $\langle p_{\tmop{rel}}, 0, 0|_{\text{bg}} W| \phi_{\tmop{res}}
\rangle$ is practically constant and does not impose any further structure on
the momentum distribution of the dissociation state.

\subsection{Connection to spectroscopy}

We now give an estimate of the matrix element
$\langle p_{\tmop{rel}}, 0, 0|_{\tmop{bg}} W| \phi_{\tmop{res}} \rangle$ in terms of
spectroscopically available quantities such as the width of the Feshbach resonance
and the background channel scattering length. A natural basis of energy eigenstates is
provided by the scattering states $| \phi^{(+)}_{\tmmathbf{p}} \rangle$, where
$[\hat{p}^2_{\tmop{rel}} / 2 \mu + V_{\tmop{bg}} (r)] | \phi^{(+)}_{\tmmathbf{p}}
\rangle = p^2 / (2 \mu) | \phi^{(+)}_{\tmmathbf{p}} \rangle$ {\cite{Kohler2006a}}.
It describes the scattering in the relative motion {\tmem{in absence}} of the confining laser
potentials, with an incoming plane wave with momentum $\tmmathbf{p}$ as boundary condition.

In order to relate $\langle p_{\tmop{rel}}, 0, 0| W| \phi_{\tmop{res}} \rangle$
to the spectroscopically available matrix element
$\langle \phi^{(+)}_{\tmmathbf{p}} |W| \phi_{\tmop{res}} \rangle$, we note that the
spatial extension of the longitudinal trap is much larger than the support of the
channel coupling $W| \phi_{\tmop{res}} \rangle$. This scale separation permits one to
approximate the effect of the longitudinal trap by a mere energy shift,
\begin{equation}
  \label{TrapEffectsEnergyShift} \langle p_{\tmop{rel}}, 0, 0|_{\tmop{bg}} W|
  \phi_{\tmop{res}} \rangle \approx \langle p_{\tmop{rel}} + p_{\text{T}}, 0,
  0|_{\text{w/oT}} W| \phi_{\tmop{res}} \rangle,
\end{equation}
where $|p_{\tmop{rel}}, 0, 0 \rangle_{\text{w/oT}}$ denotes the background
channel energy eigenstates of the relative motion {\tmem{without}} the trap
potential (without loss of generality $p_{\tmop{rel}} > 0$).
The momentum shift $p_{\text{T}}$ is related to the trap depth by
$U_{0, \text{T}} = p^2_{\text{T}} / 2 \mu$. Inserting the identity in terms
of the basis $| \phi^{(+)}_{\tmmathbf{p}} \rangle$, we can rewrite the matrix element as
\begin{multline}
  \label{UnityWithScatteringBasis} \langle p_{\tmop{rel}} + p_{\text{T}}, 0,
  0|_{\text{w/oT}} W| \phi_{\tmop{res}} \rangle\\ = \int \text{d}^3 p \langle
  p_{\tmop{rel}} + p_{\text{T}}, 0, 0|_{\text{w/oT}}
  \phi^{(+)}_{\tmmathbf{p}} \rangle \langle \phi^{(+)}_{\tmmathbf{p}} |W|
  \phi_{\tmop{res}} \rangle .
\end{multline}
The matrix element $\langle p_{\tmop{rel}} + p_{\text{T}}, 0, 0|_{\text{w/oT}}
\phi^{(+)}_{\tmmathbf{p}} \rangle$ can be evaluated for a vanishing background channel
potential, $V_{\tmop{bg}} (r) \equiv 0$, since the scattering due to $V_{\tmop{bg}}$
should not result in a substantial modification of the overlap. We thus have
\begin{align}
  \langle p_{\tmop{rel}} +& p_{\text{T}}, 0, 0|_{\text{w/oT}}
  \phi^{(+)}_{\tmmathbf{p}} \rangle\\
 & \approx  \langle p_{\tmop{rel}} +
  p_{\text{T}} |p_z \rangle \langle \varphi^{\tmop{rel}}_{0, 0} |p_x \rangle
  |p_y \rangle \nonumber \\
  & =  \delta (p_{\tmop{rel}} + p_{\text{T}} - p_z) \frac{1}{\sqrt{\hbar \omega_{\text{G}}
  \mu \pi}} \mathe^{- (p^2_x + p^2_y) / 2 \mu \hbar
  \omega_{\text{G}}} .\nonumber
\end{align}
Insertion into (\ref{UnityWithScatteringBasis}) yields
$\langle p_{\tmop{rel}} + p_{\text{T}}, 0, 0|_{\text{w/oT}} W|
\phi_{\tmop{res}} \rangle \approx \sqrt{4 \pi \mu \hbar \omega_{\text{G}}}
\langle \phi^{(+)}_0 |W| \phi_{\tmop{res}} \rangle$, where we use that
$p_{\tmop{rel}} + p_{\text{T}}$ still lies in the ultracold regime such that
$W| \phi_{\tmop{res}} \rangle$ cannot be resolved and hence
$\langle \phi^{(+)}_{\tmmathbf{p}} |W| \phi_{\tmop{res}} \rangle
\cong \langle \phi^{(+)}_0 |W| \phi_{\tmop{res}} \rangle$. This is
justified given that $p_{\tmop{rel}}$ is mainly determined by $\tilde{C}$.
With (\ref{TrapEffectsEnergyShift}) we thus obtain the desired connection to
the spectroscopically available quantity $\langle \phi^{(+)}_0 |W| \phi_{\tmop{res}} \rangle$,
\begin{equation}
   \langle p_{\tmop{rel}}, 0, 0|_{\tmop{bg}} W| \phi_{\tmop{res}} \rangle
   \approx \sqrt{4 \pi \mu \hbar \omega_{\text{G}}} \langle \phi^{(+)}_0 |W|
   \phi_{\tmop{res}} \rangle.
\end{equation}
According to Feshbach scattering theory {\cite{Kohler2006a}}
\begin{equation}
   \label{ConnectionToSpectroscopy} | \langle \phi^{(+)}_0 |W| \phi_{\tmop{res}} \rangle |^2
   = \frac{4 \pi \hbar^2}{m (2 \pi \hbar)^3} \, a_{\tmop{bg}} \, \mu_{\tmop{res}} \, \Delta B_{\tmop{res}},
\end{equation}
with $a_{\tmop{bg}}$ the background channel scattering length, $\Delta B_{\tmop{res}}$ the
resonance width, and $\mu_{\tmop{res}}$ the difference between the magnetic moments of
the Feshbach resonance state and a pair of asymptotically separated noninteracting atoms.

\subsection{Asymptotic relative motion}

Finally, let us approximate the basis states $|p_{\tmop{rel}}, 0, 0
\rangle_{\tmop{bg}}$ in the asymptotic regime $z_{\tmop{rel}} \rightarrow
\infty$, where the scattering states differ from the longitudinally
free energy eigenstates only in a scattering phase
$\varphi_{\tmop{sc}} (p_{\tmop{rel}})$,
\begin{equation}
   \langle \tmmathbf{x}_{\text{rel}} |p_{\tmop{rel}}, 0, 0 \rangle_{\tmop{bg}}
   \underset{z_{\tmop{rel}} \rightarrow \infty}{\sim}
   \mathe^{i \varphi_{\tmop{sc}} (p_{\tmop{rel}})}
   \langle z_{\text{rel}} |p_{\tmop{rel}} \rangle
   \langle \rho_{\text{rel}} | \varphi^{\tmop{rel}}_{0, 0} \rangle .
\end{equation}
The phase $\varphi_{\tmop{sc}} (p_{\tmop{rel}})$ has two
contributions, stemming from $V_{\tmop{bg}}$ and $V_{\text{T}}$.
Since we are in the ultracold regime of $V_{\tmop{bg}}$, its
contribution is a linear shift given by the background channel
scattering length $a_{\tmop{bg}}$. The contribution from $V_{\text{T}}$, on
the other hand, can be linearized due to our requirement that the energy be
sharply peaked, because the width of the energy distribution is then small
compared to the characteristic energy scale of the trap potential. The latter is
determined by the trap depth $U_{0, \text{T}}$ and is thus on the same order of
magnitude as the kinetic energy after dissociation. This situation is similar
to the scattering of a narrow wave packet with spread $\Delta E$ at a broad
resonance with width $\Gamma$, $\Delta E \ll \Gamma$ \cite{Taylor1972a}.
Confinement-induced resonances \cite{Olshanii1998a} due to the guide potential
are negligible provided the ground state size $a_{\perp}=(\hbar/m\omega_{\rm G})^{1/2}$ greatly
exceeds the background channel scattering length, $a_{\perp} \gg a_{\text{bg}}$.

As described by a linear scattering phase, the potentials thus merely
effect an overall spatial displacement of the generated dissociation
state. Physically, this shift stems from the faster propagation
of the particles in the trap region. Since we are mainly interested in the
structure of the generated dissociation state, we can safely neglect
this displacement and approximate
\begin{equation}
   \langle \tmmathbf{x}_{\text{rel}} |p_{\tmop{rel}}, 0, 0 \rangle_{\tmop{bg}}
   \underset{z_{\tmop{rel}} \rightarrow \infty}{\approx}
   \langle z_{\text{rel}} |p_{\tmop{rel}} \rangle
   \langle \rho_{\text{rel}} | \varphi^{\tmop{rel}}_{0, 0} \rangle .
\end{equation}

\subsection{Canonical dissociation state}

Putting all together, the asymptotic dissociation state reads as
\begin{align}
  \langle \tmmathbf{x}_{\text{cm}} , \tmmathbf{x}_{\text{rel}} | \Phi_{\tmop{bg}} (t) \rangle
  & \underset{t \rightarrow \infty}{\underset{|\tmmathbf{x}_{\text{rel}}| \rightarrow \infty}{\approx}} 
  \frac{1}{i \hbar}  \sqrt{4 \pi \mu \hbar \omega_{\text{G}}} \langle \phi^{(+)}_0 |W| \phi_{\tmop{res}} \rangle \nonumber \\
  &   \quad\quad\times \int^{\infty}_{- \infty} \text{d} p_{\tmop{cm}} \int^{\infty}_{- \infty} \text{d} p_{\tmop{rel}}
\nonumber\\&\quad\quad\times
  \mathe^{- i (U_{\tmop{bg}} + p^2_{\tmop{cm}} / 2 M + p^2_{\tmop{rel}} / 2 \mu) t / \hbar} \nonumber \\
  &   \quad\quad\times \tilde{C} \left( \frac{U_{\tmop{bg}} + p^2_{\tmop{cm}} / 2 M + p^2_{\tmop{rel}} / 2 \mu}{\hbar} \right)
 \nonumber \\
  &   \quad\quad\times   \langle p_{\tmop{cm}} | \psi_{\text{T}} \rangle \langle z_{\text{cm}} |p_{\tmop{cm}} \rangle \langle \rho_{\text{cm}} | \varphi^{\tmop{cm}}_{0, 0} \rangle
\nonumber\\&\quad\quad\times
  \langle z_{\text{rel}} |p_{\tmop{rel}} \rangle \langle \rho_{\text{rel}} | \varphi^{\tmop{rel}}_{0, 0} \rangle . 
\end{align}
Normalizing the spectrum with
\begin{align}
   \| \tilde{C} \|^2 =& \int^{\infty}_{- \infty} \text{d} p_{\tmop{cm}}
   \int^{\infty}_{- \infty} \text{d} p_{\tmop{rel}} | \langle p_{\tmop{cm}} | \psi_{\text{T}} \rangle |^2 \nonumber
\\&\times  | \tilde{C} \left(
   U_{\tmop{bg}} / \hbar + p^2_{\tmop{cm}} / 2 M \hbar + p^2_{\tmop{rel}} / 2
   \mu \hbar \right) |^2
\end{align}
and introducing the abbreviation
\begin{equation}
   C_{\tmop{bg}} = \frac{1}{i \hbar}  \sqrt{4 \pi \mu \hbar \omega_{\text{G}}}
   \langle \phi^{(+)}_0 |W| \phi_{\tmop{res}} \rangle \| \tilde{C} \|,
\end{equation}
we can express the dissociation state in canonical form,
\begin{equation}
  \label{DissociationStateCanonical}
  | \Phi_{\tmop{bg}} (t) \rangle \underset{t \rightarrow \infty}
  {\underset{|\tmmathbf{x}_{\text{rel}}| \rightarrow \infty}{\approx}}
  C_{\tmop{bg}} U^{(0)}_{z, t} | \Psi_z \rangle | \varphi^{\tmop{cm}}_{0, 0} \rangle
  | \varphi^{\tmop{rel}}_{0, 0} \rangle.
\end{equation}
Here $U^{(0)}_{z, t} = \exp [- i (\hat{p}^2_{\tmop{cm}} / 2 M
+ \hat{p}^2_{\tmop{rel}} / 2 \mu + U_{\tmop{bg}}) t / \hbar]$ is the free
longitudinal time evolution operator, and the longitudinal state component
$|\psi_{z}\rangle$ is defined by the momentum representation
\begin{align}
  \label{DissociationStateMomentumRepresentation} \langle p_{\tmop{cm}} |
  \langle p_{\tmop{rel}} | \Psi_z \rangle =& \frac{\tilde{C} \left(
  U_{\tmop{bg}} / \hbar + p^2_{\tmop{cm}} / 2 M \hbar + p^2_{\tmop{rel}} / 2
  \mu \hbar \right)}{\| \tilde{C} \|} \nonumber\\&\times\langle p_{\tmop{cm}} | \psi_{\text{T}}
  \rangle .
\end{align}
The dissociation probability is given by $|C_{\tmop{bg}} |^2$, which can
be expressed in terms of the above mentioned spectroscopic quantities using
(\ref{ConnectionToSpectroscopy}),
\begin{equation}
  \label{DissociationProbability} |C_{\tmop{bg}} |^2 = \frac{\omega_{\text{G}}
  a_{\tmop{bg}} \mu_{\tmop{res}} \Delta B_{\tmop{res}}}{\pi \hbar^2} \|
  \tilde{C} \|^2 .
\end{equation}
It is mainly controlled by the applied magnetic field pulse which determines
$\| \tilde{C} \|^2$. We now focus on choices of the resonance width
$\Delta B_{\tmop{res}}$ and the magnetic field pulse such that the dissociation 
probability is on the order of a few percent.

Taking Eqs.~(\ref{DissociationStateCanonical}),
(\ref{DissociationStateMomentumRepresentation}) and
(\ref{DissociationProbability}) together, we have found the desired expression
for the asymptotic dissociation state. It is characterized by the trap
geometry and, most importantly, by the Fourier transform of the closed-channel
amplitude $\tilde{C} (\omega)$, which is in turn determined by the applied
magnetic field pulse sequence. In order to answer what kind of
states can be generated, we thus have to determine the dynamics of the closed
channel amplitude $C (t)$.

\section{\label{ClosedChannelAmplitudeDynamics}closed-channel amplitude dynamics}

In the previous section we found that the momentum representation
(\ref{DissociationStateMomentumRepresentation}) of the asymptotic dissociation
state is mainly determined by the Fourier transform $\tilde{C} (\omega)$
of the closed-channel amplitude. In order to determine the generated
state for a given magnetic field pulse sequence, we thus have to determine
the dynamics of the closed-channel amplitude $C (t)$ as determined by
equation (\ref{ClosedChannelDecoupled}).

\subsection{Separation of decay and driving dynamics}

Let us rewrite the integral equation (\ref{ClosedChannelDecoupled}) as
\begin{equation}
  \label{ClosedChannelDynamics} \left[ i \hbar \partial_t - E_{\tmop{res}} (B
  (t)) - U_{\tmop{cl}} \right] C (t) = \int^t_{- \infty} \text{d} t' f (t -
  t') C (t'),
\end{equation}
with the kernel
\begin{eqnarray}
  f (\tau) & = & \frac{1}{i\hbar}\langle \psi_{\text{T}} | \langle \varphi^{\tmop{cm}}_{0,
  0} | \langle \phi_{\tmop{res}} |W U_{\tmop{bg}} (\tau) W| \psi_{\text{T}}
  \rangle | \varphi^{\tmop{cm}}_{0, 0} \rangle | \phi_{\tmop{res}} \rangle 
  \label{MemoryKernel} \nonumber \\
  & = & \frac{1}{i \hbar}  \sum_E \mathe^{- i E \tau / \hbar} | \langle
  E|_{\tmop{bg}} W| \psi_{\text{T}} \rangle | \varphi^{\tmop{cm}}_{0, 0}
  \rangle | \phi_{\tmop{res}} \rangle |^2, \nonumber
\\&&
\end{eqnarray}
where we used the decomposition of the background channel time evolution
operator (\ref{EvolutionOperatorInEnergyBasis}). The right-hand side of Eq.
(\ref{ClosedChannelDynamics}) describes the effect of the coupling between the
channels on the closed-channel amplitude. It is quadratic in the
interchannel coupling $W$ and hence, for a weak coupling and sufficiently
short dissociation windows, its effect is expected to yield a small correction
to the unperturbed dynamics given by the left-hand side. Physically, we expect it to
describe the decay of the closed-channel amplitude due to the escaping wave
packet in the background channel. The kernel (\ref{MemoryKernel}) may be viewed
as the time-dependent overlap between the ``initial state,''
$W| \psi_{\text{T}} \rangle | \varphi^{\tmop{cm}}_{0, 0} \rangle |
\phi_{\tmop{res}} \rangle$, and its evolved version $U_{\tmop{bg}} (\tau)
W| \psi_{\text{T}} \rangle | \varphi^{\tmop{cm}}_{0, 0} \rangle |
\phi_{\tmop{res}} \rangle$, which vanishes at large times due to the unbounded propagation
in the background channel. It does not depend on the external magnetic field $B (t)$.

The kernel (\ref{MemoryKernel}) is expected to drop off on a microscopic
(``memory'') time scale $t_{\text{m}}$, which can be roughly estimated from
the spatial width $\Delta x$ of the closed-channel bound state
$| \phi_{\tmop{res}} \rangle$. Denoting the corresponding momentum
uncertainty by $\Delta p$, one obtains the drop-off time scale
$t_{\text{m}} = m \Delta x^2 / \hbar$ from $\Delta p \, t_{\text{m}} / m = \Delta x$ and the
uncertainty relation. The spatial width of the closed-channel bound state
$| \phi_{\tmop{res}} \rangle$ is on the order of the closed-channel scattering
length, with typical values on the order of \ $\Delta x \approx 100 \, a_0$.
Taking the mass of lithium atoms (${}^6\tmop{Li}$) one thus
gets the estimate $t_{\text{m}} \approx 10 \, \text{ns}$, which should be compared to
the inverse decay rate of the resonance, which is much greater in our case. Given this shortness of $t_{\text{m}}$,
one might consider taking the limit $t_{\text{m}} \rightarrow 0$, which is equivalent to
setting $f (\tau) \propto \delta (\tau)$, but it will become clear later in this article that
this approximation is too crude and cannot even qualitatively account
for the correct decay behavior.

In order to separate the anticipated decay from the unitary dynamics due to the
left-hand side, we switch over to a ``comoving frame'' defined by
\begin{equation}
  \label{ClosedChannelDynamicsSeparation} C (t) = C_0 (t) D (t),
\end{equation}
where the uncoupled closed-channel amplitude $C_0 (t)$ follows by definition
from
\begin{equation}
  \label{UncoupledDynamics} \left[ i \hbar \partial_t - E_{\tmop{res}} (B (t))
  - U_{\tmop{cl}} \right] C_0 (t) = 0,
\end{equation}
which implies
\begin{equation}
  \label{UncoupledDynamicsSolution} C_0 (t) = C_0 (t_0) \, \text{exp} \left(- i
  \int^t_{t_0} \text{d} t' [E_{\tmop{res}} (B (t')) + U_{\tmop{cl}}] / \hbar \right) .
\end{equation}
Applying the ansatz (\ref{ClosedChannelDynamicsSeparation}) on (\ref{ClosedChannelDynamics})
and using (\ref{UncoupledDynamics}) and (\ref{UncoupledDynamicsSolution}), one finds that
the evolution of the coupling dynamics is governed by
\begin{align}
  \label{CouplingDynamics} i \hbar \partial_t D (t) =& \int^t_{- \infty}
  \text{d} t' D (t') f (t - t') \\&\times\text{exp}\left(i \int^t_{t'} \text{d} t''
  [E_{\tmop{res}} (B (t'')) + U_{\tmop{cl}}] / \hbar \right) .\nonumber
\end{align}
Since $D (t)$ is driven only by the coupling between the two
channels, we expect it to vary slowly for sufficiently small interchannel
coupling $W$, such that it can be considered constant to good approximation on
the time scale $t_{\text{m}}$ of nonvanishing kernel $f (t - t')$. This allows us to
pull $D (t)$ out of the integral, leading to
\[ \partial_t D (t) = \alpha (t) D (t), \]
with the (in general complex) coupling coefficient
\begin{align}
  \label{CouplingCoefficient} \alpha (t) = &\frac{1}{i \hbar} \int^t_{- \infty}
  \text{d} t' f (t - t') \nonumber\\&\times \text{exp}\left(i \int^t_{t'} \text{d} t'' [E_{\tmop{res}} (B
  (t'')) + U_{\tmop{cl}}] / \hbar \right) .
\end{align}
By writing
\begin{equation}
  \label{CouplingCoefficientRealImaginary} \alpha (t) = - \Gamma (t) / 2 - i
  \Delta E (t) / \hbar,
\end{equation}
we make explicit that the real and imaginary parts of $\alpha (t)$ describe the
decay rate $\Gamma (t)$ and an energy shift $\Delta E (t)$, respectively, as
induced by the coupling between the two channels.

\subsection{Decay dynamics}

One can evaluate the coupling coefficient (\ref{CouplingCoefficient}) further
in the case of sufficiently smooth steering of the magnetic field, such
that the resonance energy varies slowly on the scale of the drop-off time
$t_{\text{m}}$,
\begin{equation}
  \frac{\text{d}}{\text{d} t} E_{\tmop{res}} (t) \, t_{\text{m}} \ll \frac{\hbar}{t_{\text{m}}} .
\end{equation}
In the vicinity of the resonance $B_{\tmop{res}}$ one can linearize,
\begin{equation}
  \label{EnergyMagneticFieldRelation} E_{\tmop{res}} (t) = \mu_{\tmop{res}} (B
  (t) - B_{\tmop{res}}) ,
\end{equation}
leading to
\begin{equation} \label{SlowMagneticField}
  \frac{\text{d}}{\text{d} t} B (t) \ll \frac{\hbar}{t^2_{\text{m}} \mu_{\tmop{res}}} .
\end{equation}
This assumption allows one to approximate the integral in the exponent of
(\ref{CouplingCoefficient}) as
\begin{equation}
  \label{SlowMagneticFieldApproximation} \int^t_{t'} \text{d} t''
  [E_{\tmop{res}} (B (t'')) + U_{\tmop{cl}}] / \hbar \approx [E_{\tmop{res}}
  (B (t)) + U_{\tmop{cl}}] (t - t') / \hbar ,
\end{equation}
such that we can rewrite the coupling coefficient (\ref{CouplingCoefficient}) as
\begin{equation}
   \alpha (t) = \frac{1}{i \hbar} \int^{\infty}_{- \infty} \text{d} t' \Theta
   (t - t') f (t - t') \, \mathe^{i \, [E_{\tmop{res}} (B (t)) + U_{\tmop{cl}}] \,
   (t - t') / \hbar} .
\end{equation}
This can be read as the Fourier transform of the product of $f (\tau)$
and the Heaviside step function $\Theta (\tau)$, implying
\begin{align}
   \alpha (t) =& - \frac{i}{2 \hbar}  \tilde{f} \left( \frac{E_{\tmop{res}} (B
   (t)) + U_{\tmop{cl}}}{\hbar} \right)\\& + \frac{1}{2 \pi \hbar} \, \mathcal{P}
   \int^{\infty}_{- \infty} \frac{\text{d} \omega}{\omega} \,
   \tilde{f} \left( \frac{E_{\tmop{res}} (B (t)) + U_{\tmop{cl}}}{\hbar} - \omega \right) ,\nonumber
\end{align}
where $\mathcal{P}$ denotes the Cauchy principal value.
Making use of (\ref{MemoryKernel}), the Fourier transform
$\tilde{f} (\omega) = \int^{\infty}_{- \infty} \text{d} t \, \mathe^{i \omega t} f (t)$
of the kernel reads
\begin{equation}
   \tilde{f} (\omega) = \frac{2 \pi}{i \hbar}  \sum_E \delta (\omega - E /
   \hbar) | \langle E|_{\tmop{bg}} W| \psi_{\text{T}} \rangle |
   \varphi^{\tmop{cm}}_{0, 0} \rangle | \phi_{\tmop{res}} \rangle |^2 .
\end{equation}
We thus find the decay rate according to
(\ref{CouplingCoefficientRealImaginary}) to be given by
\begin{align}
  \label{DecayRate} \Gamma (t) =& \frac{2 \pi}{\hbar}  \sum_E \delta \left(
  E_{\tmop{res}} (B (t)) + U_{\tmop{cl}} - E \right)\nonumber\\&\times | \langle E|_{\tmop{bg}}
  W| \psi_{\text{T}} \rangle | \varphi^{\tmop{cm}}_{0, 0} \rangle |
  \phi_{\tmop{res}} \rangle |^2,
\end{align}
and the coupling-induced energy shift by
\begin{align}
  \label{EnergyShift} \Delta E (t) = &\sum_E \mathcal{P} \left(
  \frac{1}{E_{\tmop{res}} (B (t)) + U_{\tmop{cl}} - E} \right)\nonumber\\&\times | \langle
  E|_{\tmop{bg}} W| \psi_{\text{T}} \rangle | \varphi^{\tmop{cm}}_{0, 0}
  \rangle | \phi_{\tmop{res}} \rangle |^2 .
\end{align}
Equation (\ref{DecayRate}) shows that a nonvanishing decay rate is obtained only
when $E_{\tmop{res}} (t) + U_{\tmop{cl}}$ matches a background channel energy
eigenvalue. In particular, the gap in the spectrum between the dissociation
threshold and the highest excited bound state explains why decay occurs only
when the resonance energy lingers above the continuum threshold. (We always
stay off-tuned from bound states of $V_{\tmop{bg}}$.) This also explains why the
naive approximation for the kernel, $f (\tau) \approx f_0 \, \delta (\tau)$,
is not applicable; since the Fourier transform of $\delta (\tau)$
is constant, it cannot distinguish energies above and below the continuum
threshold and thus predicts an unphysical decay below the threshold.

Note that (\ref{DecayRate}), which coincides for constant magnetic field with the decay rate of the corresponding
Feshbach scattering resonance,
may also be viewed as a generalized version of Fermi's Golden rule, where the decay
rate is determined by the {\tmem{instantaneous}} resonance energy $E_{\tmop{res}} (B (t))$,
which in turn is externally controlled via the magnetic field $B(t)$. This coincidence
is not accidental, of course, since the limit of a slowly varying magnetic field
(\ref{SlowMagneticField}) admits both interpretations. The condition (\ref{SlowMagneticField})
quantifies the applicability of this approximation.

Let us now use the results of the preceeding section by specifying the energy
eigenbasis according to (\ref{AsymptoticQuantumNumbers}). Assuming again that
the resonance state energy $E_{\tmop{res}} (B (t))$ only sweeps over energies in
the vicinity of the background channel continuum threshold, remaining in the ultracold
energy regime, below the first excited transversal state and off-tuned from the
highest bound state of $V_{\rm bg}$, we can write
\begin{align}
  \Delta E (t)  = & | \langle p_{\tmop{rel}} = 0,n_{\text{G}}^{\text{rel}} = 0,
  m_{\text{G}}^{\text{rel}} = 0|_{\tmop{bg}} W| \phi_{\tmop{res}} \rangle |^2
\nonumber\\&\times
  \int^{\infty}_{- \infty} \text{d} p_{\tmop{cm}} \int^{\infty}_{- \infty} \text{d} p_{\tmop{rel}} \;| \langle   p_{\tmop{cm}} | \psi_{\text{T}} \rangle |^2 \\
    & \times \, \mathcal{P} \left( \frac{1}{E_{\tmop{res}} (B (t)) + U_{\tmop{cl}} - U_{\tmop{bg}} -
  \frac{p^2_{\tmop{cm}}}{ 2 M} -\frac{ p^2_{\tmop{rel}} }{ 2 \mu}} \right)  \nonumber
\end{align}
and
\begin{align}
  \Gamma (t)  = & \frac{2 \pi}{\hbar} | \langle p_{\tmop{rel}} = 0,n_{\text{G}}^{\text{rel}} = 0,
  m_{\text{G}}^{\text{rel}} = 0|_{\tmop{bg}} W| \phi_{\tmop{res}} \rangle |^2\nonumber\\&
 \times \int^{\infty}_{- \infty} \text{d} p_{\tmop{cm}}  \int^{\infty}_{-\infty} \text{d} p_{\tmop{rel}} \;  | \langle p_{\tmop{cm}}\psi_{\text{T}} \rangle |^2\\
  &   \times \, \delta (E_{\tmop{res}} (B (t)) + U_{\tmop{cl}} - U_{\tmop{bg}} - p^2_{\tmop{cm}} / 2 M
  - p^2_{\tmop{rel}} / 2 \mu)  . \nonumber
\end{align}
We have omitted transitions into bound states of the
longitudinal trap; they are negligible given the pulse sweeps
sufficiently fast over the corresponding energies. Moreover, one can arrange the dissociation pulse $B(t)$ such that the offset of the
resonance state energy from the background channel continuum threshold greatly exceeds the
trap state momentum uncertainty $\sigma_{p, \text{T}}$ for most of the time,
$E_{\tmop{res}} (B (t)) + U_{\tmop{cl}} - U_{\tmop{bg}} \gg \sigma_{p, \text{T}}^2/2 M$.
In that case the integrals are dominated by $p_{\text{rel}} \gg p_{\text{cm}}$,
and we can approximate
\begin{equation}
   \text{$| \langle p_{\tmop{cm}} | \psi_{\text{T}} \rangle |^2 \approx
   \delta (p_{\tmop{cm}})$} .
\end{equation}
This yields a vanishing energy shift, $\Delta E (t) = 0$, since the principal value
integration cancels. For the decay rate, on the other hand, we find
\begin{eqnarray}
\label{DecayRateEvaluated}
  \Gamma (t) & = & \frac{2 \omega_{\text{G}} a_{\tmop{bg}} \mu_{\tmop{res}} \Delta B_{\tmop{res}}}{\hbar}\\&&\times
  \sqrt{\frac{2 \mu}{E_{\tmop{res}} (B (t)) - 2 U_{0, \text{T}} + \hbar \omega_{\text{T}} / 2
  - \hbar \omega_{\text{G}}}} \nonumber \\
  & & \times \Theta (E_{\tmop{res}} (B (t)) - 2 U_{0, \text{T}} + \hbar \omega_{\text{T}} / 2
  - \hbar \omega_{\text{G}}), \nonumber
\end{eqnarray}
where we substituted $| \langle {p_{\tmop{rel}} = 0,n_{\text{G}}^{\text{rel}} = 0,
m_{\text{G}}^{\text{rel}} = 0|_{\tmop{bg}} }$ $ W| \phi_{\tmop{res}} \rangle |^2
= \omega_{\text{G}} a_{\tmop{bg}} \mu_{\tmop{res}} \Delta B_{\tmop{res}}
/ \pi$, $U_{\tmop{bg}} = -2 U_{0, \text{G}} + 2 \hbar \omega_{\text{G}}$ and
$U_{\tmop{cl}} = -2 U_{0, \text{T}} + \hbar \omega_{\text{T}} / 2 - 2 U_{0, \text{G}}
+ \hbar \omega_{\text{G}}$. As expected, we find that the decay rate
(\ref{DecayRateEvaluated}) is nonvanishing only for magnetic field values that lift
the resonance state above the (longitudinal) background channel continuum threshold.
The offset $-2 U_{0, \text{T}} + \hbar \omega_{\text{T}} / 2 - \hbar \omega_{\text{G}}$
in the step function gives the energy to be provided in addition to the free dissociation threshold; there, the trap depth to be 
overcome by both atoms is reduced by the closed-channel center of mass 
ground state energy $\hbar \omega_{\text{T}} / 2$, while the transversal relative motion, tightly bound in the
closed channel, must make the transition to the transversal ground state of the guide.
The square root pole stems from the one-dimensional state density and does not lead
to appreciable effects as long as the pulse sweeps sufficiently fast over it.

In summary we find, under appropriate conditions on the dissociation pulse, that
the decay dynamics of the closed-channel amplitude is described by
\begin{equation}
  \partial_t D (t) = - \frac{\Gamma (t)}{2} D (t),
\end{equation}
with the decay rate $\Gamma (t)$ given by (\ref{DecayRateEvaluated}).
By noting the formal solution $D (t) = D (t_0) \, \text{exp}\left(- \int^t_{t_0}
\text{d} t' \Gamma (t') / 2 \right)$, the overall closed-channel amplitude
$C (t)$ then follows from $C (t) = C_0 (t) D (t)$, with the uncoupled closed
channel amplitude $C_0 (t)$ given by (\ref{UncoupledDynamicsSolution}).

For the asymptotic dissociation state as described in Section \ref{AsymptoticDissociationState}
we ultimately need to know the Fourier transform of the closed-channel amplitude,
which is given by the convolution of the Fourier transforms of $C_0 (t)$ and $D (t)$,
\begin{equation}
  \label{ConvolutionTheorem} \tilde{C} (\omega) = \frac{1}{2 \pi}
  \int^{\infty}_{- \infty} \text{d} \bar{\omega}  \tilde{C}_0 ( \bar{\omega})
  \tilde{D} (\omega - \bar{\omega}) .
\end{equation}

In the limiting case of strong interchannel coupling or long-lasting, slowly
varying dissociation pulses, the kinetic energy distribution of the dissociated
atoms, denoted by $n \left( E \right)$, is determined by the decay dynamics $D
\left( t \right)$ alone. In this quasi-stationary situation one may take the
dissociation to occur monoenergetically, at the momentary resonance energy, by
writing $n \left( E \right) \mathrm{d} E = \Gamma \left( t \right) \left| D \left(
t \right) \right|^2 \mathrm{d} t$. For a monotonically increasing pulse energy
$E_{\tmop{res}} \left( t \right)$ the inverse $t(E)$    
exists (defining $\dot{E}_{\tmop{res}} \left( E \right) =\partial_t E_{\tmop{res}}(t(E)) $), and also the
decay rate (\ref{DecayRateEvaluated}) can be viewed as a function of energy. Since $D \left( t \right)$
decays exponentially it then follows that the energy distribution is given by
\begin{eqnarray}
  n \left( E \right) & = & - \frac{\mathrm{d}}{\mathrm{d} E} \exp \left( -
  \int_{E_0}^E \frac{\Gamma \left( E' \right)}{\dot{E}_{\tmop{res}} \left( E'
  \right)}  \mathrm{d} E'  \right) . 
\end{eqnarray}
This kind of quasi-stationary approach was used in {\cite{Mukaiyama2004a}} for
the case of a linear field sweep, $\dot{E}_{\tmop{res}} = \tmop{const}$, and the above formula
is consistent with their treatment when evaluated in the absence of confining
lasers.

On the other hand, in the case of a sudden magnetic field jump to a
constant value $B_0 + \Delta B$ above the threshold (which can be considered a
square-shaped pulse in the limit of infinite pulse duration), the Fourier
transform $\tilde{C}_0 (\omega)$ gets sharply peaked at $E_{\tmop{res}} (B_0 +
\Delta B)$, as will be shown later in this article. The function $\tilde{C} (\omega)$ then reduces to $\tilde{C} (\omega) \approx
\tilde{D} (\omega - E_{\tmop{res}} (B_0 + \Delta B) / \hbar)$, with $\tilde{D}
(\omega)$ a Lorentzian according to $D (t) \propto \exp (- \Gamma
(E_{\tmop{res}} (B_0 + \Delta B)) t/2)$, which recovers the corresponding
situation in \cite{Hanna2006a}.

However, in the following we are interested in the case where a magnetic field pulse
dissociates on average only a single molecule out of the BEC. This implies that the
decay of $D (t)$ can be neglected compared to the dynamics of $C_0(t)$, such that the Fourier
transform $\tilde{C} (\omega)$ is essentially given by $\tilde{C}_0 (\omega)$. We will therefore
focus on the uncoupled closed-channel amplitude $C_0 (t)$, from now on, and in particular
on magnetic field pulses that result in a sharply peaked momentum distribution,
as required for interferometric purposes. The following section is devoted to
magnetic field pulses that optimize $\tilde{C}_0 (\omega)$ with that respect. Equation
(\ref{ConvolutionTheorem}) shows that any non-negligible decay of the closed
channel amplitude then merely results in an undesired smearing of the spectrum.

\section{\label{OptimalPulse}Optimal magnetic field pulse}

We proceed to characterize the magnetic field pulse shape that is optimal in terms of providing wave packets with well-defined momentum in the undepleted molecular field limit. To this end, we can restrict the discussion to the case of a single dissociation pulse, since
the generalization to sequences of pulses is straightforward. In addition, as explained above, we are interested in situations where the depletion of the BEC can be neglected, such that the closed-channel
amplitude $C (t)$ is well approximated by the uncoupled component $C_0 (t)$, whose
dynamics is given by (\ref{UncoupledDynamics}) and (\ref{UncoupledDynamicsSolution}).

\subsection{Dimensionless formulation}

We start by switching to a more convenient representation. Since
the resonance state energy $E_{\tmop{res}} (B (t))$ is proportional to the magnetic
field $B (t)$ in the vicinity of the Feshbach resonance,
see (\ref{EnergyMagneticFieldRelation}), it is sufficient to
investigate the dependence on $E_{\tmop{res}} (B (t))$. We rewrite the
integrand in the exponent of (\ref{UncoupledDynamicsSolution}) as
\begin{equation}
  E_{\tmop{res}} (B (t)) + U_{\tmop{cl}} = E_0 + \Delta E P (t / T),
\end{equation}
where $P (t / T)$ describes a pulse with unit height; $E_0$ denotes the base value,
and $\Delta E$ and $T$ characterize the height and the width of the pulse,
respectively. Introducing the dimensionless energy $\varepsilon = \Delta E T / \hbar$
and the phase function
\begin{equation}
   \phi (t / T) = \int^{t / T}_{t_0 / T} \text{d} \tilde{t} P ( \tilde{t}) +
   \phi_0,
\end{equation}
we can write the Fourier transform of $C_0 (t)$ as
\begin{equation}
  \label{UncoupledSpectrum} \frac{\mathe^{- i E_0 t_0 / \hbar}}{C_0 (t_0)} 
  \tilde{C}_0 (\omega + E_0 / \hbar) = \int^{\infty}_{- \infty} \text{d} t
  \mathe^{i \omega t} \mathe^{- i \varepsilon \phi (t / T)} .
\end{equation}
Given that the pulse function $P ( \tilde{t})$ is positive and has a compact support,
the phase function $\phi (t)$ undergoes a monotonic ascent interpolating
between two constant levels. If the pulse is also symmetric,
$P (- \tilde{t}) = P ( \tilde{t})$, the constant of integration $\phi_0$ can
be chosen such that the phase function $\phi (t)$ is antisymmetric, $\phi (-
t) = - \phi (t)$. The constant prefactor on the left-hand side of (\ref{UncoupledSpectrum})
is irrelevant and will be neglected in the following.

\subsection{Gaussian magnetic field pulse}

We now ask for phase functions $\phi (t)$ that yield a well-behaved spectrum
$\tilde{C}_0 (\omega)$. A natural starting point is to consider the spectrum resulting
from a Gaussian-shaped pulse function $P (\tilde{t})$. In this case, as for most
other pulses, the integral on the right-hand side of (\ref{UncoupledSpectrum})
cannot be calculated exactly. But asymptotic expansion techniques can be applied
if we take the dimensionless energy $\varepsilon$ to be large, $\varepsilon \gg 1$.
This is a justified assumption in realistic scenarios. An estimate with
$\mu_{\tmop{res}} \approx 10^{- 2} \mu_{\text{B}}$, $\Delta B = 100 \tmop{mG}$
(where $B (t) = B_0 + \Delta B P (t / T)$) and $T = 100 \tmop{ms}$ yields
$\varepsilon = 10^3$.

\begin{figure}[tb]
  \includegraphics[width=\columnwidth]{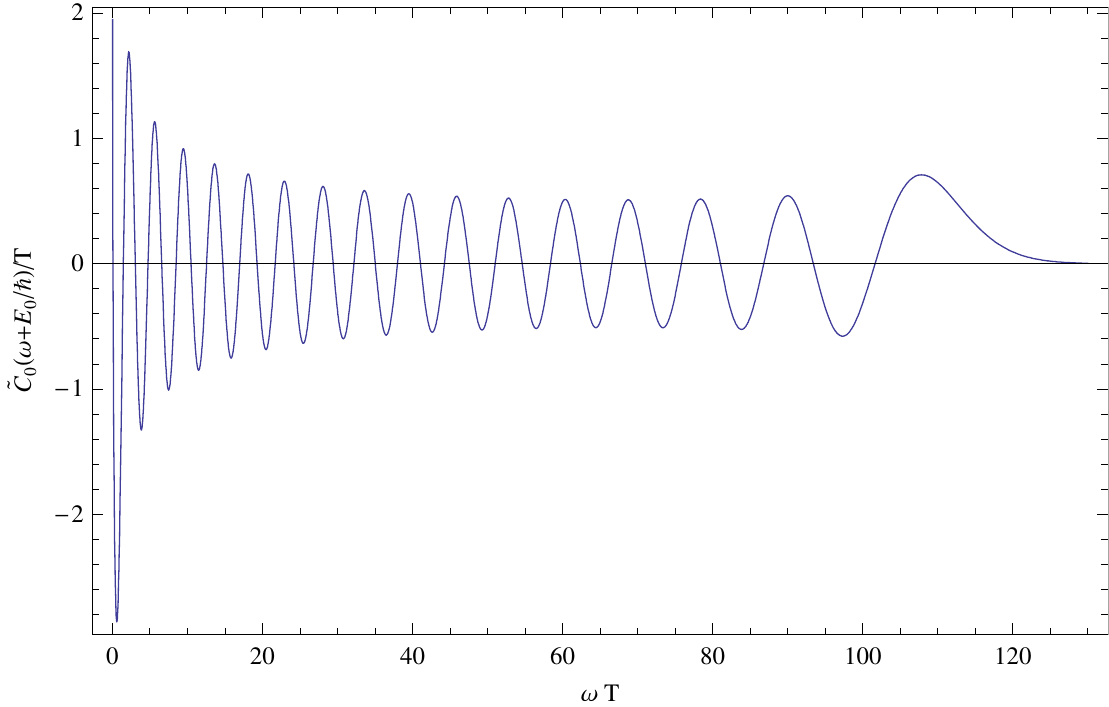}
  \caption{\label{GaussianPulseSpectrumPlot}Fourier transform $\tilde{C}_0
  (\omega)$ of the uncoupled closed-channel amplitude, here evaluated
  for a Gaussian magnetic field pulse with the dimensionless energy
  $\varepsilon = 100$. In the plot the numerically exact result is indistinguishable
  from the asymptotic evaluation as given by Eq.~(\ref{GaussianSpectrumUniform}).
  As described in Section \ref{AsymptoticDissociationState}, $\tilde{C}_0 (\omega)$
  essentially determines the momentum distribution of the asymptotic dissociation state,
  see Eq.~(\ref{DissociationStateMomentumRepresentation}). One observes that the
  spectrum does not exhibit a narrow peak-structure as one would require for
  atoms propagating with well-defined momenta. Rather, the spectrum exhibits
  contributions from all energies the pulse sweeps over. The
  oscillations stem from the interference between the contributions of the
  ascending and the descending slope of the pulse.}
\end{figure}

The Gaussian pulse leads to the error function,
\[ \phi ( \tilde{t}) = \tmop{erf} ( \tilde{t}), \]
whose analyticity admits a uniform asymptotic expansion {\cite{Bleistein1986}},
which is capable of handling stationary points both lying on the real axis
and in the complex plane. Dropping the prefactor on the left-hand
side of (\ref{UncoupledSpectrum}), the result reads
\begin{equation} \label{GaussianSpectrumUniform}
   \frac{\tilde{C}_0 (\omega + E_0 / \hbar)}{T} = 2 \pi i
   \frac{a_0}{\varepsilon^{1 / 3}} \tmop{Ai} (\varepsilon^{2 / 3} \gamma^2),
\end{equation}
with the coefficients
\[ \gamma = \left\{ \begin{array}{l}
     \begin{array}{ll}
       i \left( \frac{3}{2} \left[ \tmop{erf} \left( \alpha \right) -
       \frac{\omega T}{\varepsilon} \alpha \right] \right)^{1 / 3} & \quad\text{if\ }
       \frac{\omega T}{\varepsilon} \leq \frac{2}{\sqrt{\pi}}\\
       - \left( \frac{3}{2} \left[ \frac{\omega T}{\varepsilon} \alpha -
       \tmop{erfi} (\alpha) \right] \right)^{1 / 3} & \quad\text{if\ } \frac{\omega
       T}{\varepsilon} > \frac{2}{\sqrt{\pi}}
     \end{array}
   \end{array} \right., \]
and
\[ a_0 = \left\{ \begin{array}{l}
     \begin{array}{ll}
       - i \frac{\left( \frac{3}{2} \left[ \tmop{erf} \left( \alpha
       \right) - \omega T \alpha / \varepsilon \right] \right)^{1 /
       6}}{(\omega T \alpha / \varepsilon)^{1/2}} & \quad\text{if\ } \frac{\omega
       T}{\varepsilon} \leq  \frac{2}{\sqrt{\pi}}\\
       - i \frac{\left( \frac{3}{2} \left[ \omega T \alpha / \varepsilon
       - \tmop{erfi} (\alpha) \right] \right)^{1 / 6}}{(\omega
       T \alpha / \varepsilon)^{1/2}} & \quad\text{if\ } \frac{\omega T}{\varepsilon} >
       \frac{2}{\sqrt{\pi}}
     \end{array}
   \end{array} \right., \]
where
\[ \alpha = \left\{ \begin{array}{l}
     \begin{array}{ll}
       \sqrt{- \ln \left( \frac{\sqrt{\pi}}{2}  \frac{\omega T}{\varepsilon}
       \right)} & \quad\text{if\ } \frac{\omega T}{\varepsilon} \leq  \frac{2}{\sqrt{\pi}}\\
       \sqrt{\ln \left( \frac{\sqrt{\pi}}{2}  \frac{\omega T}{\varepsilon}
       \right)} & \quad\text{if\ } \frac{\omega T}{\varepsilon} > \frac{2}{\sqrt{\pi}}
     \end{array}
   \end{array} \right. . \]
Numerical analysis verifies excellent agreement with the exact result for large
$\varepsilon$. A plot of the corresponding spectrum for $\varepsilon = 100$
is given in Figure \ref{GaussianPulseSpectrumPlot}. It clearly does not
exhibit a well-behaved peak structure, but rather shows contributions from all
energies covered by the pulse sweep. The oscillating structure can be understood as
due to the interference between the contributions from the ascending and the
descending slope of the pulse.

\subsection{Square-shaped magnetic field pulse}

A clearer and more general insight into the relation between the
pulse and the corresponding spectrum can be obtained by retreating to the
stationary phase approximation. This comes at the cost of losing the spectrum
in the tail region where the stationary points of the exponent
in (\ref{UncoupledSpectrum}) leave the real axis. On the other hand,
also non-analytic functions can now be treated. For the sake of simplicity, we
consider symmetric pulses, $P (- \tilde{t}) = P ( \tilde{t})$. In stationary
phase approximation, we then get from (\ref{UncoupledSpectrum})
\begin{multline}
  \label{StationaryPhaseSpectrum} \frac{\tilde{C}_0 (\omega + E_0 / \hbar)}{T}
  = \sqrt{\frac{8 \pi}{\varepsilon |P' ( \tilde{t}_{\omega}) |}} \cos \left[ \omega T
  \tilde{t}_{\omega} - \varepsilon \phi ( \tilde{t}_{\omega}) + \frac{\pi}{4} \right]
  \\\text{for\ } \omega T / \varepsilon < \phi_{\max}'' ,
\end{multline}
where $\tilde{t}_{\omega}$ denotes the positive (dimensionless) stationary point, 
defined implicitly by $\phi' ( \tilde{t}_{\omega}) = \omega T / \varepsilon$. Since
$\tilde{t}_{\omega}$ is associated with the instant at which the pulse sweeps over the
frequency $\omega$, it is manifest from (\ref{StationaryPhaseSpectrum}) that
the corresponding contribution to the spectrum is the more suppressed the
faster the pulse sweeps over the corresponding energy, as one expects physically.

\begin{figure}[tb]
  \includegraphics[width=\columnwidth]{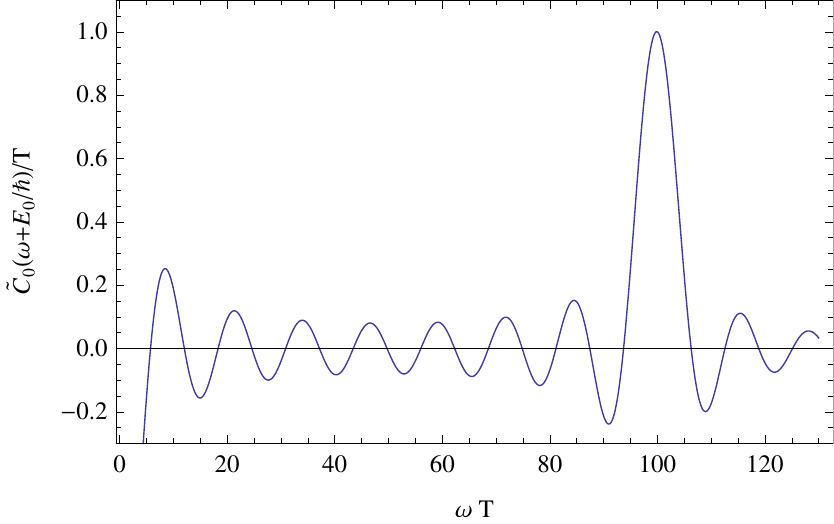}
  \caption{\label{SquarePulseSpectrumPlot}Fourier transform $\tilde{C}_0
  (\omega)$ of the uncoupled closed-channel amplitude component
  for a square-shaped magnetic field pulse with dimensionless energy
  $\varepsilon = 100$. In contrast to the Gaussian pulse spectrum, it exhibits
  a pronounced peak at $\omega T = \varepsilon$. Together with the
  connection (\ref{DissociationStateMomentumRepresentation})
  between $\tilde{C} (\omega)$ and the asymptotic dissociation state, this
  proves the capability of the scheme to provide relatively well-defined atom momenta,
  even though there is no way to tailor arbitrary dissociation states.
  In particular, sequences of such pulses then can serve to generate highly
  nonclassical motionally entangled states. (Since $E_0 < 0$ the pole at $\omega T= 0$
  does not affect the asymptotic behavior of the atoms.)}
\end{figure}

Aiming at a spectrum that is well-peaked at a specific energy, we thus must
strive for magnetic field pulses that sweep as fast as possible over the region
of undesired energies and then rest at a plateau determined by the desired energy.
The (idealized) optimal pulse with that respect is a square-shaped pulse,
\begin{equation}
  \label{SquarePulseFunction} P ( \tilde{t}) = \Theta ( \tilde{t} + 1 / 2)
  \Theta ( \tilde{t} - 1 / 2),
\end{equation}
for which (\ref{UncoupledSpectrum}) can be calculated directly, yielding
\begin{align}
  \label{SquarePulseSpectrum} \frac{\tilde{C}_0 (\omega + E_0 / \hbar)}{T} = &2
  \pi \cos \left( \frac{\varepsilon}{2} \right) \delta (\omega T) 
\nonumber\\&+
  \tmop{sinc} \left( \frac{\omega T - \varepsilon}{2} \right) \mathcal{P}
  \left( \frac{\varepsilon}{\omega T} \right) .
\end{align}
The pole at $\omega T = 0$ can be traced back to the fact that the pulse
function (\ref{SquarePulseFunction}) is nonvanishing only on a finite time
interval. For $E_0 < 0$, which corresponds to an asymptotic magnetic field in
the bound regime, this pole lies at a negative energy irrelevant for
the shape of the dissociation state, as was shown in Section
\ref{AsymptoticDissociationState}. For our purposes it is thus
sufficient to restrict the discussion to positive frequencies with
$\omega + E_0 / \hbar > 0$. A plot of (\ref{SquarePulseSpectrum}) for
$\varepsilon = 100$ is shown in Figure \ref{SquarePulseSpectrumPlot}.
As anticipated, the spectrum exhibits a pronounced peak at $\omega T = \varepsilon$,
whose width is characterized by the pulse duration $T$. According to
(\ref{DissociationStateMomentumRepresentation}), the corresponding dissociation
state therefore exhibits a narrow momentum distribution. This was used in
{\cite{Gneiting2009b}} where the two-particle momentum distribution
generated by a square pulse has been worked out for a specific experimental
scenario.

This concludes the quest for the optimal magnetic field pulse shape. As
indicated in the stationary phase calculation (\ref{StationaryPhaseSpectrum}),
by sweeping as fast as possible over undesired energies one singles out a spectrum
peaked at the desired energy. This has been verified numerically for a large
variety of pulse forms. In particular, it is now clear that smoothening
the edges of the square shape inevitably enhances the undesired ripples
in the region of the energies swept over, contradicting the naive expectation
that they stem from the non-differentiable ansatz (\ref{SquarePulseFunction})
for the pulse.

More generally, we found that the nature of Feshbach dissociation dynamics puts strong
limitations on the possible range of states that can be generated. It is in general not
possible to find a pulse shape so as to generate a specified dissociation
state. On the other hand, as was demonstrated in {\cite{Gneiting2008a}}, the
combination of a sequence of two dissociation pulses yields a promising
perspective to generate highly nonclassical motionally entangled two-atom states
with the capacity to violate a Bell inequality.

\section{\label{Conclusions}Conclusions}

We described the dissociation of Feshbach molecules as triggered by a time-varying
external magnetic field in a realistic trap and guide setting. Employing a coupled
channels formulation including the center of mass motion, we derived an analytic
expression for the asymptotic dissociation state in the laser guide. It is
essentially determined by the Fourier transform of the closed-channel amplitude $C (t)$.
The established relation between applied dissociation pulse and resulting spectrum
$\tilde{C} (\omega)$ puts strong limitations on the possibility to tailor a pulse
shape in order to obtain a desired dissociation state. However, a square-shaped
magnetic field pulse is shown to optimize the momentum distribution with respect
to its width, as it is required for subsequent interferometric processing.

The use of the Feshbach dissociation scheme as a source for entangled atom
states was demonstrated in {\cite{Gneiting2008a}}. There, a sequence of two
equal magnetic field pulses dissociates the molecules such that
each atom is delocalized into a pair of macroscopically separated, consecutive
wave packets. Their subsequent interferometric processing on each side then
constitutes a Bell test whose success requires a sharp momentum distribution
as provided by the discussed square-shaped pulses.

An alternative protocol that can do without the interferometers employs
a sequence of dissociation pulses of different heights. Given that the second pulse
is the stronger one, the corresponding wave packets propagate faster as compared
to the ones associated with the first pulse, such that time evolution alone
effects their overlap. Position measurements in the overlap region on each side
and their appropriate dichotomization may again reveal nonclassical correlations
that entail a Bell violation.

One may also discard the restriction to the dissociation of a
single pair and consider the simultaneous dissociation of a multitude of molecules
in a multipair Bell setting {\cite{Bancal2008a}}. Statistical effects due to the
indistinguishability of the particles then require a theoretical treatment beyond
the single-molecule dissociation dynamics. For this, the above derived two-body transition
amplitude (between molecular initial state and asymptotic dissociation state) could
serve as an essential building block, for example with respect to a cumulant expansion
{\cite{Kohler2002a}}.



\end{document}